\documentclass[twocolumn,twocolappendix]{aastex702}
\usepackage{bm,CJK}

\newcommand{\Alfven}{Alfv\'{e}n\ }
\newcommand{\Alfvenic}{Alfv\'{e}nic\ }

\begin{document}
\begin{CJK*}{UTF8}{ipxg}
\title{Dependence of Particle Acceleration Efficiency on Shock Velocity \\in Weakly Magnetized Electron-Ion Shocks}

\author[orcid=0000-0002-1876-5779]{Taiki Jikei (寺境太樹)}
\affiliation{Department of Astronomy and Columbia Astrophysics Laboratory, Columbia University, 538 West 120th Street, New York, NY 10027, USA}
\email[show]{t.jikei@columbia.edu} 

\author[orcid=0000-0002-5408-3046]{Daniel {Gro{\v{s}}elj}} 
\affiliation{Centre for mathematical Plasma Astrophysics, Department of Mathematics, KU Leuven, Celestijnenlaan 200B, B-3001 Leuven, Belgium}
\email{daniel.groselj@kuleuven.be}

\author[orcid=0000-0002-1227-2754]{Lorenzo Sironi}
\affiliation{Department of Astronomy and Columbia Astrophysics Laboratory, Columbia University, 538 West 120th Street, New York, NY 10027, USA}
\affiliation{Center for Computational Astrophysics, Flatiron Institute, 162 5th Avenue, New York, NY 10010, USA}
\email{lsironi@astro.columbia.edu}

\begin{abstract}
Using unprecedentedly long 2D particle-in-cell simulations, we study electron and ion acceleration in weakly magnetized quasi-parallel shocks, propagating at velocities ranging from transrelativistic to subrelativistic.
At a fixed upstream magnetic field strength, low-velocity quasi-parallel shocks are dominated by the Bell instability, whereas high-velocity shocks are dominated by the Weibel instability.
Both regimes accelerate ions with similar efficiency, with the Bell-dominated regime exhibiting faster growth in the maximum particle energy.
The electron acceleration efficiency is strongly dependent on shock velocity.
Weibel-dominated shocks have $\sim15\,\%$ of shock energy in nonthermal electrons, whereas in the Bell-dominated regime we attribute less than $\sim2\,\%$ of shock energy to nonthermal electrons.
We discuss applications of our results to the bright X-ray emission from the late-stage afterglows of gamma-ray bursts, the radio emission from fast blue optical transients, and the X-ray variability in microquasars.
\end{abstract}

\section{Introduction} \label{sec:intro}
The physics of collisionless shocks propagating at transrelativistic velocities has attracted much interest following recent observations, such as the long-term monitoring of the gamma-ray burst (GRB) afterglow of the binary neutron star merger event GW170817 \citep{Margutti2018, Hajela2019, Hajela2022}, the discovery of fast blue optical transients (FBOTs) \citep{Margutti2019, Ho2019, Ho2022}, and the detection of $\gtrsim100\,\mathrm{TeV}$ gamma rays from several microquasars by LHAASO \citep{Cao2025}.
These shocks are all considered to have transrelativistic velocities and propagate in weakly magnetized plasmas.\footnote{Precise definitions of shock velocity and magnetization will be presented in section \ref{sec:method}.}
Therefore, one may expect their particle-acceleration characteristics to be qualitatively similar. 
However, observations suggest a very different picture.

The afterglow of GW170817 shows clear signs of nonthermal X-ray emission even $\sim1000$ days after the burst \citep{Hajela2022}.
In contrast, radio emission from some FBOTs suggests that they may be powered by relativistic thermal electrons \citep{Ho2022}.
This discrepancy implies qualitatively distinct electron-acceleration physics despite only modest differences in the shock parameters.
Considering microquasars, SS 433 consistently shows bright X-ray knots \citep{Tsuji2025}, while V4641 Sgr only shows strong X-ray emission during occasional X-ray outbursts \citep{Uemura2002, Shaw2022}.
Yet, the highest-energy gamma-ray emission from V4641 Sgr exceeds that of SS 433 \citep{Cao2025}.
In this work, we aim to identify, from a plasma-microphysics perspective, the mechanisms responsible for the diverse behaviors that emerge within a relatively narrow range of shock velocities and magnetizations.

In \citet{Jikei2026}, we presented a series of particle-in-cell (PIC) simulations with a fixed shock velocity and variable magnetization, and discussed the physics of magnetic field amplification and particle acceleration.
We found that a small variation in the ambient magnetization could result in drastic changes in the magnetic field generation at the shock.
At relatively high magnetizations $(\sigma\gtrsim10^{-3})$, the shock precursor is dominated by fluctuations generated by the Bell instability \citep{Bell2004}.
In contrast, the Weibel instability \citep{Weibel1959, Fried1959} dominates at lower magnetizations $(\sigma<10^{-4})$, as expected in GRB afterglow shocks.
The differences in the magnetic field structure also led to qualitatively different particle-acceleration properties for both ions and electrons.
Here, we perform a new set of simulations with a fixed magnetization but different shock velocities to investigate the dependence of particle acceleration on shock velocity.
Although we use a fixed shock velocity for each simulation run, our results can also be used to discuss the effect of a change in shock velocity (e.g., blast wave deceleration) on dynamical timescales, which are much longer than the plasma scales of our simulations.

We find a sharp transition from the Bell-dominated regime to the Weibel-dominated regime as we increase the shock velocity at fixed magnetization, similar to what we found in \citet{Jikei2026} as we decreased the upstream magnetization at fixed shock velocity.
In the Bell-dominated regime, ions are accelerated very efficiently, but the electron acceleration efficiency is much smaller than in the Weibel regime.
The Weibel-dominated regime accelerates both ions and electrons with similar efficiency, but at a slower rate than ions in the Bell-dominated regime.
We discuss the implications of our PIC results for the emission from late-stage GRB afterglows, FBOTs, and microquasars.

\section{Method} \label{sec:method}
We perform 2D shock simulations with the PIC code OSIRIS \citep{Fonseca2002, Fonseca2013}.
The setup, which is described here for completeness, closely follows \citet{Jikei2026}.
To investigate the dependence on shock velocity, we fix the upstream magnetic field and run multiple simulations with different upstream flow velocities.
In the upstream frame, we define the magnetization parameter as
\begin{equation}
\sigma=\frac{B_0^2}{4\pi n_0 (m_{\mathrm{i}}+m_{\mathrm{e}})c^2},
\end{equation}
where $B_0$ is the amplitude of the ambient magnetic field in the upstream frame, $n_0$ is the comoving number density, and $m_s$ is the mass of particle species $s$ (i for ions and e for electrons).
As in \citet{Jikei2026}, we study a quasi-parallel shock configuration, where
the angle $\theta_B$ between  $\bm{B}_0$ and the shock normal is small.
We use $\sigma=10^{-4.5}=3.16\times10^{-5}$ and $\theta_B=20^{\circ}$.
The simulations are performed in the $x-y$ plane in the downstream frame, in which the upstream plasma drifts in the negative $x-$direction with a drift speed of $v_0$ and collides against a reflecting wall.
The upstream density in the simulation frame is $N_0=\gamma_0n_0$, and the ambient field in the simulation frame is $\bm{B}_{0|\mathrm{d}}=(B_0\cos\theta_B, \gamma_0B_0\sin\theta_B, 0)$, where $\gamma_0=1/\sqrt{1-(v_0/c)^2}$.
An in-plane ambient magnetic field configuration is chosen to capture the physics of particles and waves propagating along field lines.
We vary the upstream four-velocity in the range $u_0/c=[1/6, 4/3]$, where $u_0=\gamma_0v_0$.

The box size in the transverse $(y)$ direction is $163.84\,d_{\mathrm{i}}$, where $d_{\mathrm{\mathrm{i}}}=c/\omega_{\mathrm{pi}}$ is the ion skin depth.
We define the plasma frequency of species $s$ as $\omega_{\mathrm{p}s}=(4\pi n_0e^2/m_s)^{1/2}$, and $c$ is the speed of light.
The gyrofrequency of species $s$ is given by $\Omega_s=eB_0/m_sc$. 
Periodic boundaries are used in the $y-$direction.
Shock formation is triggered by a reflecting wall boundary located at $x=0$.
A moving injector traveling at the speed of light in the $+x$-direction continuously supplies the upstream plasma.
The cell size and the time step are $\Delta x/d_{\mathrm{e}}=0.4$ and $c\Delta t/\Delta x=0.5$, respectively.
The upstream plasma is initialized with $N_{\mathrm{ppc}}=16$ particles per cell per species, and we use cubic-spline shape functions.
The simulation ion-to-electron mass ratio is set to $m_{\mathrm{i}}/m_{\mathrm{e}}=100$ to save computational resources.
We apply four passes of binomial filtering in each direction to the electric current.
See \citet{Jikei2026} and references therein for further optimizations, which allow us to run the simulations up to $\omega_{\mathrm{pi}}t=10000$.

\section{Results} \label{sec:results}
\subsection{Shock Structure}
\begin{figure*}[htb!]
\includegraphics[width=\linewidth]{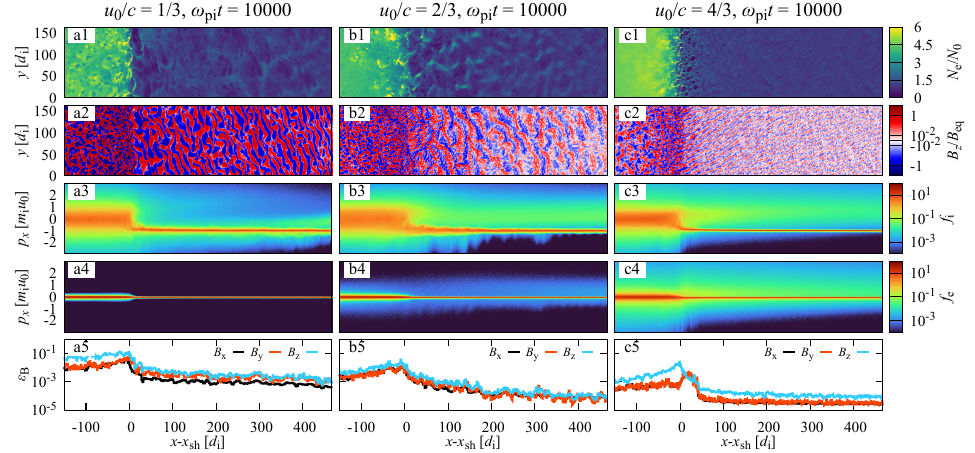}
\caption{Snapshot of the simulations taken at $\omega_{\mathrm{pi}}t=10000$. 
         Columns a, b, and c correspond to $u_0/c=1/3, 2/3$, and $4/3$, respectively.
         Row 1: electron number density $N_{\mathrm{e}}$ normalized by the far upstream value $N_0$.
         Row 2: $z$-component of the magnetic field normalized by the equipartition field (Eq.~(\ref{eq:equipart_field})).
         The colorbar is on a symmetric log scale, in which the range $[-0.01, 0.01]$ is on a linear scale.
         Rows 3 and 4: phase space density $f_s(x,p_x)$ for ions and electrons, respectively.
         Row 5 shows the normalized magnetic field energy density  (Eq.~(\ref{eq:eps_b})) for each component.
         Black, orange, and light blue correspond to $B_x, B_y$, and $B_z$, respectively.
         }
\label{fig:shock}
\end{figure*}

Let us first compare the shock structures for different upstream flow speeds $u_0=\gamma_0v_0$.
Figure~\ref{fig:shock} shows snapshots taken at $\omega_{\mathrm{pi}}t=10000$.
Columns a, b, and c correspond to $u_0/c=1/3, 2/3,$ and $4/3$, respectively.
The $u_0/c=4/3$ run was previously discussed in \citet{Jikei2026}.
Here, we define the location of the shock front $x_{\mathrm{sh}}$ as the position at which, starting from far upstream, the $y$-averaged ion density first exceeds $2.5\,N_0$.
Row 1 shows the electron number density $N_{\mathrm{e}}/N_0$.
All three runs exhibit strong shocks with a similar compression ratio of $\sim4$, comparable to the jump conditions for an adiabatic index of $\sim5/3$ \citep{Landau1987}.
The density fluctuations in both the upstream and downstream regions differ noticeably among the three runs.
The shock with the slowest upstream speed, $u_0/c=1/3$ (a1), exhibits the largest cavity-like density structures in the upstream region, as well as the strongest downstream density fluctuations.
The shock with the fastest speed, $u_0/c=4/3$ (c1), exhibits the smallest-scale upstream density fluctuations, and its downstream region is closest to uniform density.
Somewhat intermediate characteristics are seen for $u_0/c=2/3$ (b1).
These density structures are closely coupled to the magnetic structure, as we now describe.

Row 2 shows the $z$-component of the magnetic field, normalized by the equipartition field
\begin{equation} \label{eq:equipart_field}
B_{\mathrm{eq}}=\sqrt{4\pi n_0\gamma_0(\gamma_0-1)(m_{\mathrm{i}}+m_{\mathrm{e}})c^2}.
\end{equation}
The $z$-component is chosen because it best characterizes different shock regimes \citep[see][for details]{Jikei2026}.
The two runs on the left (a2, b2) show magnetic-field fluctuations with a wavenumber predominantly parallel to the ambient field.
These waves exhibit circular polarization, a characteristic of the Bell instability \citep{Bell2004, Amato2009, Caprioli2014a, Caprioli2014b, Crumley2019}.
The case $u_0/c=1/3$ (a2) has stronger magnetic field fluctuations (when normalized to $B_{\mathrm{eq}}$) than the case $u_0/c=2/3$ (b2).
We show in subsection \ref{subsec:self-regulation} that the runs with $u_0/c=1/3$ and $2/3$ lie in different regimes of the Bell instability at the time shown in Figure~\ref{fig:shock}, leading to different magnetic field saturation strengths.
In particular, the shock with $u_0/c=2/3$ operates in the high-current regime of the Bell instability \citep{Weidl2019, Lichko2025}. 
In contrast, by the time shown in the figure, the $u_0/c=1/3$ shock has transitioned into the low-current regime, where magnetic field amplification is more efficient.
The rightmost $u_0/c=4/3$ run shows a magnetic field structure with a wavenumber predominantly perpendicular to the ambient field, and its typical spatial scale is noticeably smaller than in the other two runs.
This can be attributed to magnetic field generation by the Weibel instability \citep{Spitkovsky2008a, Spitkovsky2008b, Kato2007, Kato2008, Groselj2024}.
Row 5 shows the $y$-averaged component-wise magnetic field energy, defined as 
\begin{equation} \label{eq:eps_b}
\varepsilon_{B,j}=B_{j}^2/B_{\mathrm{eq}}^2.
\end{equation}
where $j=x,y,z$.
We see that the field generated by the Bell instability (a5, b5) has $\varepsilon_{B,y}\sim\varepsilon_{B,z}$, which confirms that the magnetic fluctuations are circularly polarized. 
In contrast, the Weibel-dominated case (c5) has $\varepsilon_{B,z}\gg\varepsilon_{B,y}$, consistent with the typical structure of Weibel fields in a 2D simulation of unmagnetized or weakly magnetized shocks.
Note that this is a simplified argument: it neglects the effect of a finite magnetic-field obliquity, $\theta_B\neq0$, in the Bell-dominated case, and the effect of a finite ambient field, $B_0\neq0$, in the Weibel-dominated case.
We will now show how these magnetic field structures affect the particle distributions.

Rows 3 and 4 show the particle phase-space distribution functions $f_s(x,p_x)$ for ions and electrons, respectively.
Looking at the returning ions with momenta $p_x>0$ in the upstream region $(x>x_{\mathrm{sh}})$, we notice that the Bell-dominated  $u_0/c=1/3$ case (a3) has a much smaller fraction of returning ions, compared to the Weibel-dominated  $u_0/c=4/3$ case (c3).
This indicates that a self-regulation process is at work.
In \citet{Jikei2026}, we showed that Bell-dominated shocks can adjust the returning current to a level favorable for efficient magnetic field amplification by lowering the fraction of returning ions.
Recall that in that paper we used a fixed shock velocity and varied the upstream magnetization.
The result presented here implies that a similar argument can be made for a fixed magnetization but at different shock velocities.
We will further elaborate on the self-regulation process in subsection \ref{subsec:self-regulation}.

Electrons exhibit an even stronger dependence on shock velocity.
In the downstream $(x<x_{\mathrm{sh}})$ of the Bell-dominated $u_0/c=1/3$ shock (a4), we see a very small fraction of high-energy electrons with $p_x>m_{\mathrm{i}}u_0$, compared to ions in the same run or electrons in other runs.
In contrast, there is a comparable number of high-energy electrons as high-energy ions for the Weibel-dominated $u_0/c=4/3$ run (c4).
These trends are qualitatively consistent with our previous work, which focused on a mildly relativistic regime with $u_0/c=4/3$ \citep{Jikei2026}.
At a fixed velocity $u_0/c=4/3$, shocks with higher magnetizations ($\sigma\gtrsim10^{-3}$) were Bell-dominated, and shocks with lower magnetizations ($\sigma\lesssim10^{-4}$) were Weibel-dominated.
The former had noticeably fewer high-energy electrons downstream, while the latter had comparable numbers of high-energy electrons and ions.
However, a quantitative assessment of the fraction of nonthermal particles requires extra care, since we are comparing shocks with different velocities and thus different degrees of relativistic effects.
We investigate the particle acceleration of ions and electrons in subsection \ref{subsec:acceleration}, and we combine the results of this work with our previous paper \citep{Jikei2026} in subsection \ref{subsec:summary}.

In addition to the three runs reported in Figure~\ref{fig:shock}, we have performed simulations with $u_0/c=1/6$ and $1$, which are not shown here.
The former shows characteristics similar to those of the $u_0/c=1/3$ run, and the latter resembles the $u_0/c=4/3$ shock.
These runs will be included in the analysis in the following subsections.
 
\subsection{Cosmic-Ray Self Regulation} \label{subsec:self-regulation}
As theory suggests \citep{Weidl2019, Lichko2025}, and was confirmed for fixed shock speed and varying magnetizations \citep{Jikei2026}, the magnetic amplification efficiency in the shock upstream strongly depends on the level of cosmic-ray current. 
Here, we define the cosmic-ray current $J_{\mathrm{cr}}$ as the current carried by the upstream ions that have had $p_x>0$ at least once (all PIC ions have  $p_x<0$ at initialization).
In the upstream frame, the linear growth rate of the Bell instability is proportional to the projection of the cosmic-ray current along the ambient field \citep{Bell2004, Amato2009}:
\begin{equation} \label{eq:bell_growth}
\frac{\Gamma_{\mathrm{Bell}}}{\omega_{\mathrm{pi}}}=\frac{\eta}{2}.
\end{equation}
Here, we have defined the normalized current $\eta=J_{\mathrm{cr}}/n_0ec$.
For efficient magnetic field amplification, which we define as $\delta B/B_0\gg1$, the linear growth rate (Eq.~(\ref{eq:bell_growth})) needs to be smaller than the gyro-frequency of background protons, i.e., 
\begin{equation} \label{eq:current_raw}
\Gamma_{\mathrm{Bell}}<\Omega_{\mathrm{i}}.
\end{equation}
We can define the maximum current that satisfies Eq.~(\ref{eq:current_raw}) as the critical current 
\begin{equation} \label{eq:crit_current}
\eta_{\mathrm{crit}}=2\sigma^{1/2}.
\end{equation}
Above this current, which defines the transition to the high-current regime \citep{Weidl2019}, the ambient field strength determines the saturation amplitude of the Bell instability.
\citet{Lichko2025} finds that the saturated field strength in the high-current regime is $\delta B\sim5\,B_0$, which can be smaller than the magnetic field generated by the Weibel instability in weakly magnetized environments \citep{Jikei2026}.

\begin{figure}[htb!]
\includegraphics[width=\linewidth]{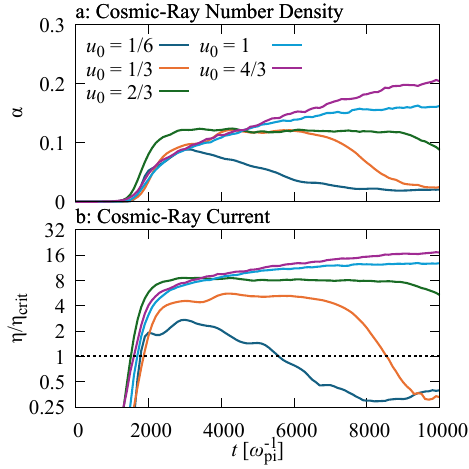}
\caption{Time evolution of the cosmic-ray ion number density and current.
         a: number density ratio $\alpha$ in the upstream frame.
         b: upstream current normalized by the critical current, $\eta/\eta_{\mathrm{crit}}$.
         Both are computed in $(x-x_{\mathrm{sh}})/d_{\mathrm{i}}=[200, 300]$.
         Dark teal, orange, dark green, turquoise, and purple represent $u_0/c=1/6, 1/3, 2/3, 1$ and $4/3$, respectively.
         Note that the vertical axis of panel b is on a logarithmic scale.}
\label{fig:current}
\end{figure}

Figure~\ref{fig:current} shows the time evolution of the current of cosmic-ray ions in the near upstream region $(x-x_{\mathrm{sh}})/d_{\mathrm{i}}=[200, 300]$.
Panel a shows the cosmic-ray-to-background ion number density ratio, $\alpha=N_{\mathrm{cr}|\mathrm{u}}/n_0$,  measured in the upstream frame.
Weibel-dominated shocks with $u_0/c=1, 4/3$, as well as the early stages of runs with $u_0/c=1/6, 1/3$ (before the onset of the Bell instability), typically have $\alpha\gtrsim0.1$.
Since $\alpha$ is defined in the upstream frame, more relativistic cases $(u_0/c=1, 4/3)$ could exhibit larger values of $\alpha$ as a consequence of frame transformations, even if the reflection probability in the shock frame is the same.
In contrast,  Bell-dominated shocks regulate the number of returning ions to $\alpha\ll0.1$ after the Bell instability enters the nonlinear stage (at $\omega_{\mathrm{pi}}t\sim7000$ for $u_0/c=1/6$ and $\omega_{\mathrm{pi}}t\sim9000$ for $u_0/c=1/3$).

Panel b shows the time evolution of the normalized upstream ion current projected along the ambient magnetic field, $\eta/\eta_{\mathrm{crit}}=J_{\mathrm{cr|u}}/(2\sigma^{1/2}n_0ec)$.
The $u_0/c=1/6$ and $1/3$ shocks self-regulate the current to $\eta/\eta_{\mathrm{crit}}\lesssim0.5$, resulting in a low-current Bell regime, while the Weibel-dominated $u_0/c=1$ and $4/3$ shocks show no signs of current self-regulation within the timespan covered by our simulations.
The intermediate $u_0/c=2/3$ case is in the middle of the transition from high to low current at the end of our simulation.
In this run, we see the cosmic-ray current starting to drop at $\omega_{\mathrm{pi}}t\gtrsim9000$.
Although our simulations do not guarantee that shocks with high cosmic-ray currents, $u_0/c=1$ and $4/3$, will indefinitely maintain $\eta/\eta_{\mathrm{crit}}>1$, two series of supplementary simulations suggest that the current remains sufficiently high for these shocks to stay Weibel-dominated.
In Appendix \ref{app:twocompbell}, we present periodic-box simulations in which the cosmic rays consist of both ions and electrons. These simulations show that the relativistic returning electrons present in Weibel-dominated shocks, as shown in Figure~\ref{fig:shock}c4, suppress the growth and saturation of the Bell instability and make the upstream environment more favorable to the Weibel instability.
We also ran the $u_0/c=4/3$ shock simulation far beyond $\omega_{\mathrm{pi}}t=10000$, albeit with a narrower $ y$-box (see Appendix \ref{app:longterm}). Until $\omega_{\mathrm{pi}}t=10000$, the results of this narrow box agree with those of our fiducial run. At later times, the simulation with a narrower box confirms that the shock remains Weibel-dominated up to at least $\omega_{\mathrm{pi}}t=25000$. 

In summary, we find that the competition between the Bell and Weibel instabilities is closely coupled to the self-regulation of the returning ion current.
The Bell-dominated regime adjusts the returning ion current to $\eta/\eta_{\mathrm{crit}}\lesssim0.5$, while  Weibel-dominated shocks feature a high cosmic-ray current $(\eta/\eta_{\mathrm{crit}}>1)$ and a fraction of reflected particles that does not decrease over time $(\alpha\gtrsim0.1)$.
For $\sigma=10^{-4.5}$, the transition between the two regimes occurs around $u_0/c\sim2/3$.

\subsection{Particle Acceleration} \label{subsec:acceleration}
\begin{figure}[htb!]
\includegraphics[width=\linewidth]{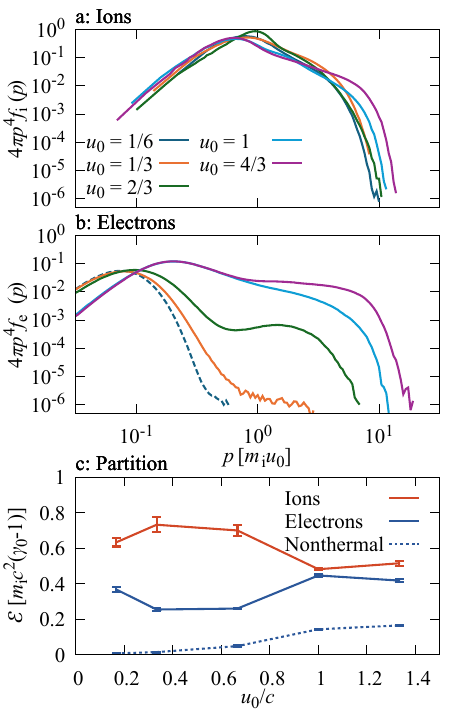}
\caption{Particle momentum spectrum $4\pi p^4f(p)$ at the end of the simulations, $\omega_{\mathrm{pi}}t=10000$, in the near downstream, $(x-x_{\mathrm{sh}})/d_{\mathrm{sh}}=[-100, -50]$.
         a: ions. b: electrons.
         The color code is the same as in Figure~\ref{fig:current}.
         The bottom panel shows the downstream kinetic energy partition, defined as the average particle energy normalized by the upstream ion energy.
         Red and blue correspond to ions and electrons.
         The blue dotted line represents the nonthermal electron energy fraction, defined as the fraction of total energy contained in electrons with $p>3\,p_{\mathrm{peak}}$.
         Error bars are calculated from temporal variability in the time range   $\omega_{\mathrm{pi}}t=[7000, 10000]$.
}
\label{fig:spectrum}
\end{figure}

Figure~\ref{fig:spectrum} shows the momentum spectrum of particles located in the near downstream region $(x-x_{\mathrm{sh}})/d_{\mathrm{i}}=[-100, -50]$, taken at $\omega_{\mathrm{pi}}t=10000$.
The distribution functions are normalized such that
\[
\int_0^{\infty}4\pi p^2f_s(p)dp=\int_1^{\infty}f_s(\gamma)d\gamma=1.
\]
Here we plot $4\pi p^4f_s(p)$, so that a nonthermal power-law tail of the form $f(p)\propto p^{-4}$, as expected in the test particle limit for strong nonrelativistic shocks, would appear flat \citep[e.g.,][]{Caprioli2014a, Park2015}.
The particle momentum on the horizontal axis is normalized to $m_{\mathrm{i}}u_0$, which allows us to easily compare runs with different shock velocities.
Panel a shows the ion spectrum.
In all runs, the thermal ion peak is located near $p/m_{\mathrm{i}}u_0\sim1$, and the nonthermal tail extends up to $p/m_{\mathrm{i}}u_0\sim10$ (but its upper cutoff steadily increases over time; see below).
Note that the growth rate of the maximum ion energy depends on the upstream magnetic field structure, which determines the scattering regime, as we discuss below.

Panel b shows the electron spectra, which exhibit a stark dependence on shock velocity.
There appear to be two distinct values for the electron thermal momentum.
The Bell-dominated shocks with $u_0/c=1/6, 1/3$ and $2/3$ have an electron thermal  peak at $p/m_{\mathrm{i}}u_0\sim0.07$.
The thermal peak of the Weibel-dominated runs ($u_0/c=1$ and $4/3$) is located at a noticeably larger momentum, $p/m_{\mathrm{i}}u_0\sim0.2$.
The normalization of the electron nonthermal tail, measured at the point beyond the thermal peak where $4\pi p^4 f_{\mathrm{e}}(p)$ begins to flatten, shows an even stronger dependence on $u_0$.
The nonthermal tail of the Weibel-dominated run with $u_0/c=4/3$ emerges just below the thermal peak.
In contrast, the Bell-dominated shock with $u_0/c=1/3$ has a nonthermal tail with a normalization orders of magnitude below the thermal peak.
This is consistent with \citet{Jikei2026}, who found that Weibel-dominated shocks channel more energy into post-shock nonthermal electrons than Bell-dominated shocks. We attributed this result to the fact that the Bell instability generates large-amplitude transverse magnetic fields with wavelengths larger than the Larmor radius of thermal electrons, making the shock effectively perpendicular for electrons and thus suppressing their injection.
The $u_0/c=2/3$ case exhibits intermediate behavior.
However, this may not be the final state since the shock structure is still evolving (see Figure~\ref{fig:current}).
We have verified the numerical convergence of the electron nonthermal level for the $u_0/c=1/3$ and $2/3$ cases using narrower-box runs, $L_y=10.24\,d_{\mathrm{i}}$, with four times more particles per cell, $N_{\mathrm{ppc}}=64$. 
We do not further discuss the nonthermal electron tail for $u_0/c=1/6$, as it may not be numerically converged \citep{Kato2013, May2014}.

Panel c of Figure~\ref{fig:spectrum} shows the downstream energy partition, which we define as
\begin{equation}
\mathcal{E}_s=\frac{m_sc^2(\left<\gamma\right>_s-1)}{m_{\mathrm{i}}c^2(\gamma_0-1)},
\end{equation}
where $\left<\gamma\right>_s$ is the average Lorentz factor of species $s$:
\begin{equation}
\left<\gamma\right>_s=\frac{\int_1^{\infty}\gamma f_s(\gamma)d\gamma}{\int_1^{\infty}f_s(\gamma)d\gamma}.
\end{equation}
We also calculate the electron nonthermal energy fraction, defined as the fraction of energy contained in electrons with momentum three times larger than the momentum of the thermal peak, $p>3\,p_{\mathrm{peak}}$, where $p_{\mathrm{peak}}=m_{\mathrm{e}}c\sqrt{\gamma_{\mathrm{peak}}^2-1}$ and  $\gamma_{\mathrm{peak}}$ is the electron Lorentz factor at the peak of $(\gamma-1)f_{\mathrm{e}}(\gamma)$.
We use the same near-downstream region $(x-x_{\mathrm{sh}})/d_{\mathrm{i}}=[-100, -50]$, and we average in time over the interval $\omega_{\mathrm{pi}}t=[7000, 10000]$.
Error bars are calculated from the temporal variability during this period.
Although the energy of the incoming flow is dominated by ions, in the downstream region a fraction $0.25-0.45$ of the total energy is carried by electrons  ($0.5$ would correspond to equipartition between ions and electrons).
These values are consistent with previous simulations \citep[e.g.,][]{Vanthieghem2024, Tran2024}. 
The nonmonotonic dependence of the electron-to-ion temperature ratio on shock velocity at low velocities is beyond the scope of this work but remains an important topic for future studies \citep{Raymond2023}.

Shocks dominated by the Weibel instability ($u_0/c=1$ and $4/3$) convert a fraction $\sim0.15$ of the total energy ($\sim0.3$ of the post-shock electron energy) to downstream nonthermal electrons.
In contrast, shocks dominated by the Bell instability convert much less than $0.1$ of the shock energy into nonthermal electrons.
For the $u_0/c=1/3$ shock, downstream nonthermal electrons contain a fraction $\sim 0.015$ of the total energy, or $\sim 0.06$ of the post-shock electron energy. The $u_0/c=1/6$ shock has an even smaller nonthermal fraction, although we were unable to assess numerical convergence for this case.

\begin{figure}[htb!]
\includegraphics[width=\linewidth]{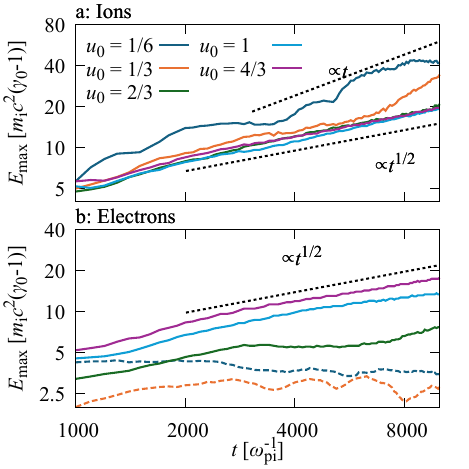}
\caption{Time evolution of the maximum energy (Eq.~(\ref{eq:maximum_energy})), normalized by $m_{\mathrm{i}}c^2(\gamma_0-1)$.
a: ions, b: electrons.
The color code is the same as in Figure~\ref{fig:current}.
Note that both horizontal and vertical axes are on a log scale.
The black dotted lines indicate power-law scalings.}
\label{fig:maximum_energy}
\end{figure}

The rate at which the maximum particle energy evolves is a crucial aspect of the physics of particle acceleration.
The maximum energy of species $s$ is
\begin{equation} \label{eq:maximum_energy}
E_{\mathrm{max},s}=m_sc^2(\gamma_{\mathrm{max},s}-1),
\end{equation}
where $\gamma_{\mathrm{max},s}$ is the Lorentz factor at which $(\gamma-1)f_s(\gamma)$ drops below $10^{-5}$ of its peak value \citep{Sironi2013, Groselj2024}.
Figure~\ref{fig:maximum_energy} shows the time evolution of $E_{\mathrm{max},s}(t)$ calculated in the downstream region $(x-x_{\mathrm{sh}})/d_{\mathrm{i}}=[-100, -50]$ (the same region as used in Figure~\ref{fig:spectrum}).
As shown by panel a, the maximum ion energy steadily increases for all shock velocities.
The Weibel-dominated cases with $u_0/c=1$ and $4/3$, and the $u_0/c=2/3$ run in which Bell modes are present but have low amplitudes, display the scaling  $E_{\mathrm{max, i}}\propto t^{1/2}$ expected for acceleration in small-scale fields \citep{Sironi2013, Jikei2026}.
The Bell-dominated shocks exhibit a different acceleration rate.
The $u_0/c=1/3$ run (orange) transitions from $E_{\mathrm{max, i}}\propto t^{1/2}$ to $E_{\mathrm{max, i}}\propto t$ (or even a bit faster) at late times, when the Bell instability becomes dominant (see Figure~\ref{fig:current}).
The $u_0/c=1/6$ shock, in which the Bell instability is dominant since early times, shows consistently fast acceleration with $E_{\mathrm{max, i}}\propto t$, as expected for Bell-dominated shocks \citep{Gargate2012, Stockem2012}.
The non-steady growth of the maximum ion energy in this case (not a straight line in the plot) can be attributed to shock reformation \citep[e.g.,][]{Balogh2013, Burgess2015}, which is more pronounced in lower-Mach-number shocks.

\subsection{Summary of the Results} \label{subsec:summary}
In this work, we compared a series of 2D shock simulations with a fixed magnetization $\sigma=10^{-4.5}$ and varying shock velocity in the range $u_0/c=[1/6, 4/3]$.
Shocks with low velocities ($u_0/c=1/6$ and $1/3$) are Bell-dominated, whereas high-velocity shocks ($u_0/c=1$ and $4/3$) are Weibel-dominated, with intermediate behavior seen for $u_0/c=2/3$.
Both Bell-dominated and Weibel-dominated shocks accelerate ions with similar efficiency (Figure~\ref{fig:spectrum}a), but Bell-dominated shocks display faster acceleration rates (Figure~\ref{fig:maximum_energy}a).
These trends are consistent with our earlier study \citep{Jikei2026}, in which we fixed the velocity to $u_0/c=4/3$ and varied the magnetization in the range $\sigma=[10^{-4.5}, 10^{-3}]$.
In that work, we observed a transition between Bell-dominated and Weibel-dominated regimes when the magnetization dropped below $\sigma\sim10^{-3.5}$.

We can synthesize our new results and the conclusions of \citet{Jikei2026} by introducing the \Alfvenic Mach number.
For $\sigma\ll1$, which is our regime of interest, the \Alfven velocity $V_{\mathrm{A}}$ is related to the magnetization as $V_{\mathrm{A}}/c\sim\sqrt{\sigma}$.
If we relate the normalized cosmic-ray current and the normalized cosmic-ray number density (see subsection \ref{subsec:self-regulation}) via $\eta\sim\alpha v_0/c$, which assumes that upstream cosmic-ray ions drift with a mean speed of $\sim v_0$ (equivalently, they are roughly isotropic in the downstream frame), we can rephrase the low-current condition in Eq.~(\ref{eq:crit_current}) as
\begin{equation} \label{eq:current_ma}
\alpha<\alpha_{\mathrm{crit}}=2M_{\mathrm{A}}^{-1},
\end{equation}
where $M_{\mathrm{A}}=v_0/V_{\mathrm{A}}$ is the \Alfvenic Mach number.
In both this paper, where we fix the magnetization, as well as in \citet{Jikei2026}, where we fixed the flow speed, we see a transition between Bell-dominated and Weibel-dominated regimes at $M_{\mathrm{A}}\sim100$.
Shocks with $M_{\mathrm{A}}<100$ tend to reduce $\alpha$ to satisfy Eq.~(\ref{eq:current_ma}), resulting in a Bell-dominated shock.
At $M_{\mathrm{A}}>100$ no self-regulation to the low-current Bell regime is observed, and the shock stays in the Weibel regime with $\alpha\gtrsim0.1$.
In the nonrelativistic limit, the efficiency of magnetic field generation by the Weibel instability may be suppressed \citep{Jikei2024b, Law2025}, possibly resulting in a higher value of the critical $M_{\mathrm{A}}$ separating Bell and Weibel regimes.
Simulations with smaller velocity $v_0\ll0.1\,c$ and magnetization $\sigma\ll10^{-6}$ are needed to clarify whether the transition at $M_{\mathrm{A}}\sim100$ still holds for nonrelativistic shocks (see subsection \ref{subsec:future_work} for further discussion).

Another important finding of this work is that the energy fraction carried by nonthermal electrons depends strongly on the shock velocity.
The electron nonthermal tail in Weibel-dominated shocks with $u_0/c\gtrsim1$ lies less than one order of magnitude below the peak, whereas for Bell-dominated $u_0/c\lesssim1/3$ shocks the electron nonthermal tail is about five orders of magnitude below the peak.
The strong dependence of the electron acceleration efficiency on shock velocity in the transrelativistic regime can be important in several astrophysical sources, as we 
further discuss in section \ref{sec:discussion}.

\section{Discussion} \label{sec:discussion}
\subsection{Late Time Afterglows of GRBs}
We found that the critical \Alfvenic Mach number is $\sim100$, above which the shock is Weibel-dominated and accelerates electrons efficiently (Figure~\ref{fig:spectrum}b).
For GW170817, assuming that the off-axis jet Lorentz factor is $\Gamma\sim4$ at 100 days after the burst \citep{Mooley2018}, the characteristic Lorentz factor after 1000 days is $\sim1.6$ for a Blandford-McKee scaling \citep{Blandford1976}, which assumes no jet spreading, and $\sim1.1$ for an exponential jet spreading \citep{Rhoads1999}, respectively \citep{Hajela2022}.
The corresponding velocities are $v\sim0.8\,c$ and $0.4\,c$.
In an interstellar-like medium, the \Alfven velocity is 
\begin{equation}
V_{\mathrm{A}}\sim2\times10^{-5}\,c\,\left(\frac{B}{3\,\mathrm{\mu G}}\right)\left(\frac{n}{1\,\mathrm{cm}^{-3}}\right)^{-1/2},
\label{eq:vA_ISM}
\end{equation}
resulting in $M_A\sim2\times10^4$ for $v=0.4\,c$, which is much larger than the critical Mach number $M_A\sim 100$ identified in this work.
This is consistent with the detection of bright X-ray emission throughout the afterglow evolution of GW170817 \citep{Margutti2018, Hajela2019, Hajela2022}.
The brightness of the nonthermal emission may drop abruptly when the \Alfvenic Mach number falls below $\sim100$, i.e., the velocity becomes $\sim2\times10^{-3}\,c$.
Starting with $v=0.4\,c$ at $t=1000$ days and assuming the Taylor-von Neumann-Sedov scaling \citep{Landau1987}, we have
\begin{equation}
v(t)=0.4\,c\,\left(\frac{t}{1000\,\mathrm{days}}\right)^{-2/5},
\end{equation}
which drops below $\sim100\,V_{\mathrm{A}}$ only at $t\sim6\times10^{8}$ days. Therefore, we expect the late-time afterglow shock to remain efficient at accelerating electrons over any observationally accessible timescale.
The transition to inefficient electron acceleration 
would be occurring earlier if the shock were to propagate in a higher-magnetization medium with large $V_{\mathrm{A}}$ (e.g., a magnetized stellar wind).
Note that in realistic decelerating shocks, particles that were accelerated earlier, when the shock velocity was higher, may be reaccelerated in the later-stage, lower-velocity shock.
Thus, there remains the possibility that even $M_{\mathrm{A}}<100$ Bell-dominated shocks can keep accelerating the highest energy electrons, which were pre-accelerated during an earlier Weibel-dominated stage.

\subsection{Radio Emission From FBOTs}
In contrast to late-stage GRB afterglows, which show clear signs of nonthermal emission throughout their evolution, observations suggest that the radio emission from FBOTs may be powered by relativistic thermal electrons.
In particular, \citet{Ho2022} show excellent agreement between the observational data and a thermal electron model for AT2020xnd.
Our results imply that the energy fraction of nonthermal electrons could drop significantly at $M_{\mathrm{A}}\lesssim100$.
Recall that the \Alfvenic Mach number can be written as
\begin{equation}
M_{\mathrm{A}}=100\left(\frac{v}{0.1\,c}\right)\left(\frac{\sigma}{10^{-6}}\right)^{-1/2}.
\end{equation}
Although GRB afterglows propagating in a $\sigma\sim10^{-9}$ medium will likely have $M_{\mathrm{A}}\gg100$, some FBOTs may have \Alfvenic Mach numbers lower than 100 if the magnetization of the surrounding medium is higher.

Our findings may also be used for advancing theoretical models of transrelativistic transients.
For example, \citet{Margalit2021} \citep[extended to include relativistic effects in][]{Margalit2024} proposed a model to infer the shock velocity from the peaks of the observed spectral energy distribution (SED).
Their model assumes a fixed nonthermal energy fraction for all shock velocities.
More accurate modeling will be possible by incorporating the dependence of electron acceleration efficiency on shock velocity as found in this work.
Such theoretical refinements will become increasingly important as new observations and improved data-analysis methods yield more accurate SEDs.

\subsection{X-Ray Variability in Microquasars}
Microquasars have gained significant attention in recent years, following the detection of ultra-high-energy gamma rays by LHAASO \citep{Cao2025}.
It is quite puzzling that microquasars, e.g., SS 433 and V4641 Sgr, behave very differently from each other, even though they are all $\sim100\,\mathrm{TeV}$ gamma-ray sources harboring transrelativistic jets.
SS 433 has been consistently bright in X-ray knots over the past $\sim20$ years \citep[e.g.,][]{Tsuji2025}.
In contrast, V4641 Sgr emits bright X-rays only during outbursts \citep{Uemura2002, Shaw2022}.
The maximum energy of observed photons from V4641 Sgr is $\sim800\,\mathrm{TeV}$, noticeably larger than other microquasars, including SS 433 \citep{Cao2025}.

Here, we propose a possible explanation for such a variety of signatures, assuming that particles are accelerated at transrelativistic shocks.
We postulate that SS 433 is usually in the Weibel-dominated state and V4641 Sgr is usually in the Bell-dominated state.
This corresponds to assuming that SS 433 has either a weaker upstream magnetization \citep{Jikei2026} or a higher shock velocity (this work).
This assumption is consistent with the bright X-ray emission observed from SS 433, which would be powered by nonthermal electrons.
If V4641 Sgr is in the Bell-dominated regime during its quiescent state, our results imply that it may not accelerate electrons efficiently enough to power bright X-ray emission,  in agreement with the observations.
X-ray outbursts may be attributed to an increase in shock velocity, which transitions V4641 Sgr into a Weibel-dominated state, thereby increasing the electron acceleration efficiency and emitting bright X-rays.

Regarding the maximum energy of gamma rays, we propose a scenario in which they have a hadronic origin.
Both Bell-dominated and Weibel-dominated shocks accelerate ions with similar efficiencies (Figure~\ref{fig:spectrum}a).
Therefore, we argue that transrelativistic shocks in microquasars can generate high-energy protons, regardless of the magnetic field regime.
However, we note that the growth of the maximum energy is faster for the Bell-dominated regime $(E_{\mathrm{max}}\propto t)$, compared to the Weibel-dominated regime $(E_{\mathrm{max}}\propto t^{1/2})$, see Figure~\ref{fig:maximum_energy}a.
These arguments are consistent with the fact that V4641 Sgr, which we assume is usually in the Bell-dominated regime, is the source of the highest-energy gamma rays among the microquasars observed by LHAASO.

\subsection{Future Work} \label{subsec:future_work}
The parameter space we investigated in this work is roughly $\sigma<10^{-4}$ and $v_0>0.2\,c$.
Here, we briefly comment on the physics expected in neighboring regimes.
First, transrelativistic shocks with higher magnetization $10^{-3}\lesssim\sigma\lesssim0.1$ have been discussed in previous works \citep{Crumley2019, Jikei2026}.
Shocks in this regime are Bell-dominated; thus, they are promising candidates for extreme ion acceleration in extragalactic jets \citep{Cerutti2023, Globus2025}.
PIC simulations of transrelativistic shocks with $\sigma\gtrsim0.1$ have mostly been limited to perpendicular shocks \citep[e.g.,][]{Ligorini2021A, Ligorini2021B}.
The quasi-parallel regime is underinvestigated, and it may lead to efficient particle acceleration.

In this work, we focused on shock velocities larger than $v_0/c=1/6$.
High-Mach-number $(M_{\mathrm{A}}\gtrsim100)$ nonrelativistic $(v_0<0.1\,c)$ shocks are important in the context of young supernova remnants (SNRs) and laboratory laser experiments.
Multidimensional PIC simulations modeling SNR or planetary shocks \citep{Matsumoto2015, Bohdan2021, Jikei2024a} and laboratory laser experiments \citep{Fox2013, Huntington2015, Fiuza2020} suggest that the Weibel instability plays an important role in this regime. On the other hand, 1D PIC simulations, which usually employ lower Mach numbers $(M_{\mathrm{A}}<50)$, show evidence of particle acceleration associated with the Bell instability \citep[e.g.,][]{Park2015}. All of these studies have limitations. Multidimensional PIC simulations of SNR-like shocks have primarily focused on quasi-perpendicular field configurations, while laboratory experiments typically operate in regimes where the background field is negligible. Also, 1D PIC simulations of parallel shocks cannot capture the physics of the Weibel instability. Thus, a systematic study of these shock regimes using multidimensional PIC simulations of quasi-parallel configurations is still lacking.

Finally, \citet{Gupta2024} show that, in the nonrelativistic limit ($v_0 \ll 0.1\,c$), the electron acceleration efficiency is higher for slower shocks. When combined with our results (Figure~\ref{fig:spectrum}), this raises the possibility that the electron acceleration efficiency is minimized for shocks with velocities around $v_0\sim0.1\,c$. Future multidimensional PIC simulations of high-Mach-number quasi-parallel shocks will be essential for clarifying the shock structure and particle acceleration in this regime.

\begin{acknowledgments}
We thank Siddhartha Gupta, Ben Margalit, Raffaella Margutti, and Kaya Mori for fruitful discussions.
T.J. and L.S. are supported by grants from the Simons Foundation (MP-SCMPS-0000147).
T.J. is also supported by NASA ATP 80NSSC24K1826.
D.G. is supported by the Research Foundation--Flanders (FWO) Senior Postdoctoral Fellowship 12B1424N.
L.S. was also supported by the NSF grant PHY2409223, by the Multimessenger Plasma Physics Center (MPPC, NSF grant PHY2206609), and by the DoE Early Career Award DE-SC0023015. 
This research used computational resources of the National Energy Research Scientific Computing Center (NERSC), a Department of Energy User Facility, under NERSC award FES-ERCAP0037296; the supercomputer of ACCMS, Kyoto University, through the HPCI System Research Project (Project ID: hp250112); and the Flatiron Institute. 
We acknowledge the OSIRIS Consortium, consisting of UCLA and IST (Lisbon, Portugal), for the use of OSIRIS and for providing access to the OSIRIS framework.
\end{acknowledgments}

\appendix
\section{Streaming Instability Driven by Electron-Ion Cosmic Rays} \label{app:twocompbell}
In the main text, we showed that, at a fixed magnetization $(\sigma=10^{-4.5})$, the Bell instability is the dominant upstream instability for $u_0/c\lesssim2/3$.
In contrast, faster $u_0/c\gtrsim1$ shocks remain Weibel-dominated (for the $u_0/c=4/3$ case, we tested this conclusion up to $\omega_{\mathrm{pi}}t=25000$, see Appendix \ref{app:longterm}).
This implies that some mechanism suppresses the Bell instability at high shock velocities.
However, previous studies based on a pure ion cosmic-ray beam did not exhibit such a dependence on shock velocity.
In the low-current regime $\eta/\eta_{\mathrm{crit}}<1$, \citet{Zacharegkas2024} showed, with analytical considerations and hybrid simulations, that the saturation level of the Bell mode is
\begin{equation} \label{eq:sat_low_current}
\frac{\delta B^2}{8\pi\beta_{\mathrm{d}}T^{01}_{\mathrm{b,i}}}\sim\frac{1}{4},
\end{equation}
where $\delta B$ is the strength of the fluctuating magnetic field, $T^{\mu\nu}_{\mathrm{b,i}}$ is the stress energy tensor of the beam, and $\beta_{\mathrm{d}}$ is the drift velocity of the beam relative to the background plasma.
In the high-current regime $(\eta/\eta_{\mathrm{crit}}>1$, \citealt{Weidl2019}), the saturation level is capped by the ambient field strength.
By means of PIC simulations, \citet{Lichko2025} found that in this regime
\begin{equation} \label{eq:sat_high_current}
\left(\frac{\delta B}{B_0}\right)^2\sim25.
\end{equation}
For fixed beam momentum flux and fixed background magnetic field, these expressions contain no explicit dependence on the shock velocity, in contrast to the velocity dependence found in our simulations.\footnote{We investigated the saturation level of the high-current Bell mode in \citet{Jikei2026}.
          We used $\beta_{\mathrm{d}}=0.8$, which is larger than in most of the runs by \citet{Lichko2025} and in the simulations of this appendix.
          With $\beta_{\mathrm{d}}=0.8$, we found a somewhat larger saturation level, $(\delta B/B_0)^2\sim100$, across different magnetizations and beam energies.
          This implies a dependence on the beam Lorentz factor, which may be of interest when considering ultra-relativistic beams.
          For the current work, the crucial point is that the saturation of the Bell instability in the high-current regime is capped by the ambient field strength and is typically lower than that in the low-current regime.
          Therefore, we will refrain from discussing further the factor of $\sim4$ difference between our earlier results \citep{Jikei2026} and \citet{Lichko2025}.
          }

In this appendix, we show that such a velocity dependence would be introduced if we consider the effect of returning electrons.
We perform 1D periodic-box simulations using the open-source code SMILEI \citep{Derouillat2018}.
We work in the rest frame of the background plasma with number density $n_0$, and set the ambient field $B_0$ in the $x$-direction (along the 1D box),
An electron-ion beam with a proper number density $n_{\mathrm{b},s}$ drifts in the $x$-direction, along the ambient field, with a drift velocity $v_{\mathrm{d}}$.
Note that the simulation frame density of beam particles is $N_{\mathrm{b},s}=\gamma_{\mathrm{d}}n_{\mathrm{b},s}$.
The beam Lorentz factor is $\gamma_{\mathrm{b}}=1/\sqrt{1-\beta_{\mathrm{d}}^2}$ with $\beta_{\mathrm{d}}=v_{\mathrm{d}}/c$.
We use a Maxwell-J{\"u}ttner distribution \citep{Synge1957} for beam ions, with a temperature defined as:
\begin{equation} \label{eq:temperature}
3\Theta_{\mathrm{b,i}}+\frac{K_1(1/\Theta_{\mathrm{b,i}})}{K_2(1/\Theta_{\mathrm{b,i}})}=\gamma_{\mathrm{d}},
\end{equation}
with $\Theta_{\mathrm{b,i}}=k_{\mathrm{B}}T_{\mathrm{b,i}}/m_{\mathrm{i}}c^2$.
This means that the average thermal Lorentz factor in the beam rest frame equals the beam drift Lorentz factor in the simulation frame.\footnote{Eq.~(\ref{eq:temperature}) asymptotes to $\Theta_{\mathrm{b,i}}\sim\gamma_{\mathrm{d}}/3$ in the ultrarelativistic limit $(\gamma_{\mathrm{d}}\to\infty)$ and $\Theta_{\mathrm{b,i}}\sim2(\gamma_{\mathrm{d}}-1)/3$ in the nonrelativistic limit $[(\gamma_{\mathrm{d}}-1)\to 0]$.
Here, we use the full expression because we are in the transrelativistic regime.}
For simplicity, we set the number density and temperature of the electron beam to be the same as the ion beam: $N_{\mathrm{b,e}}=N_{\mathrm{b,i}}$ and $T_{\mathrm{b,e}}=T_{\mathrm{b,i}}$.
We fix the magnetization to $\sigma=10^{-4}$ and compare different drift velocities in the range of $\beta_{\mathrm{d}}=[0.2, 0.7]$.
We use a box size of $L_x=153.6\,d_{\mathrm{i}}$ with grid size of $\Delta x=0.1\,d_{\mathrm{e}}$.
The time step is $c\Delta t/\Delta x=0.99$.
The simulation mass ratio is $m_{\mathrm{i}}/m_{\mathrm{e}}=100$. 
We use $N_{\mathrm{ppc}}=1024$ particles per cell per species, with quartic spline shape functions.
Since the characteristics of the Bell instability differ between low-current and high-current regimes \citep{Weidl2019, Jikei2026}, we conduct separate numerical experiments for each regime.

\begin{figure*}[htb!]
\includegraphics[width=\linewidth]{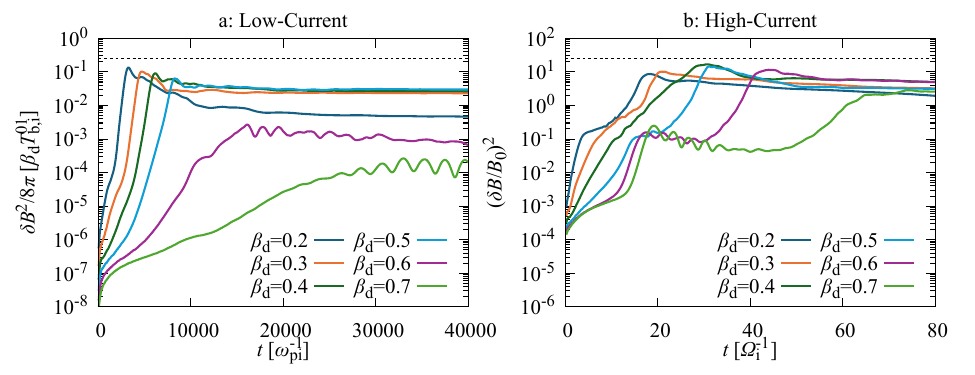}
\caption{Time evolution of the Bell-amplified magnetic field energy for an electron-ion beam.
         a: results of the low-current regime, in which time is normalized by the ion plasma frequency and the magnetic field energy is normalized by the momentum flux of the ion beam.
         b: results of the high-current regime, in which time is normalized by the ion gyrofrequency, and the magnetic field energy is normalized by the ambient field energy.
         Dark teal, orange, dark green, turquoise, purple, and green curves represent $\beta_{\mathrm{d}}=0.2, 0.3, 0.4, 0.5, 0.6$, and $0.7$, respectively.
         The black dotted lines are the theoretical estimates of the saturation level for a pure ion beam (Eq.~(\ref{eq:sat_low_current}) for panel a and Eq.~(\ref{eq:sat_high_current}) for panel b).}
\label{fig:two_comp_bell}
\end{figure*}
Figure~\ref{fig:two_comp_bell}a shows the time evolution of the magnetic field energy amplified by the Bell instability in the low-current regime.
We fix the normalized ion-only current $\eta_{\mathrm{b,i}}=\beta_{\mathrm{d}}N_{\mathrm{b,i}}/n_0=0.5\,\eta_{\mathrm{crit}}$.
In other words, we decrease the beam number density as we increase the drift velocity to maintain the same normalized current, which is set to half of the critical value.
Note that the net current of the beam, i.e., the sum of ion and electron currents, is zero.
The horizontal axis is time, normalized by the background ion plasma frequency.
The linear growth rate for a pure ion beam is $\Gamma_{\mathrm{Bell}}=\eta_{\mathrm{b,i}}/2$, which is independent of $\beta_{\mathrm{d}}$ for our choice of a fixed $\eta_{\mathrm{b,i}}$.
The vertical axis is the magnetic field energy normalized by the ion momentum flux, which in this case reads
\begin{equation}
\beta_{\mathrm{d}}T^{01}_{\mathrm{b,i}}=m_{\mathrm{i}}c^2N_{\mathrm{b,i}}\gamma_{\mathrm{d}}\beta^2_{\mathrm{d}}\left[4\Theta_{\mathrm{b,i}}+\frac{K_1(1/\Theta_{\mathrm{b,i}})}{K_2(1/\Theta_{\mathrm{b,i}})}\right].
\end{equation}
The four runs with the lowest drift velocities, $\beta_{\mathrm{d}}=0.2, 0.3, 0.4$, and $0.5$, show saturation levels relatively close to the value expected for a pure ion beam (Eq.~(\ref{eq:sat_low_current})).
In these cases, the relative velocity between the beam electrons and the background electrons drops nearly to zero, presumably due to the leptonic Bell instability \citep{Gupta2021}.
After the beam electrons decelerate, the beam ion current is exposed, and the background plasma effectively interacts with a pure ion beam.
These runs have different onset times for the ion-driven Bell instability due to differences in the timescale of electron deceleration, but the growth rates after electron deceleration are rather consistent.
The two high-drift-velocity runs ($\beta_{\mathrm{d}}=0.6$ and $0.7$) show a saturation level that is clearly lower than the pure-ion prediction.
Because of their highly relativistic temperatures, electrons do not decelerate prior to ions in these cases. As a result, they continue to carry a current that efficiently compensates the ion current before the onset of the ion-driven instability. This reduces the net current available to amplify the magnetic field, thereby suppressing the Bell instability.

Therefore, we find that current compensation by an ultrarelativistic electron beam can suppress the Bell instability in the low-current regime.
However, such a current compensation may have the opposite effect in the high-current regime.
If the ion current is in the high-current regime, one may speculate that efficient magnetic field amplification would be possible if---due to partial compensation by the beam electrons---the net current is in the low-current regime.
To test this, we run the same experiment, now in the high-ion-current regime.
We fix $\eta_{\mathrm{b,i}}=\beta_{\mathrm{d}}N_{\mathrm{b,i}}/n_0=4\,\eta_{\mathrm{crit}}$.
Figure~\ref{fig:two_comp_bell}b shows the results.
Note that the normalizations of time and energy have now been changed to $\Omega_{\mathrm{i}}$ and $B_0^2$, respectively, to match the expectations of \citet{Weidl2019, Lichko2025}.
For all values of drift velocity, the saturation level for electron-ion beams does not exceed the pure-ion saturation level (Eq.~(\ref{eq:sat_high_current})).
Note that the high-current Bell instability driven by a pure ion beam is less efficient than its low-current counterpart, as is apparent if the fluctuation energy $\delta B^2/8\pi$ is consistently normalized in both regimes (e.g., to the beam momentum flux $\beta_{\mathrm{d}}T^{01}_{\mathrm{b,i}}$).
The two-step growth of the curves in panel b, i.e., the fact that the magnetic field energy settles to $(\delta B/B_0)^2\sim10^{-1}$ at earlier times, before a second growth that exceeds $(\delta B/B_0)^2\sim1$ at a later stage, can be attributed to the leptonic Bell instability.
We have confirmed this by checking the polarization of the resulting (circularly-polarized) modes, which is opposite between the early phase and the secondary growth.
We refrain from providing further details on this because our main focus is on the final saturation level.

In this Appendix, we have shown that ultrarelativistic electron beams suppress the Bell instability in both low- and high-current regimes.
This is consistent with the results in the main text, in which low-velocity shocks, with fewer relativistic electrons, transition to a Bell-dominated state, while higher-velocity shocks, with relativistic electrons, are Weibel-dominated.

\section{Longer Term Evolution with Narrower Box} \label{app:longterm}
\begin{figure}[htb!]
\includegraphics[width=\linewidth]{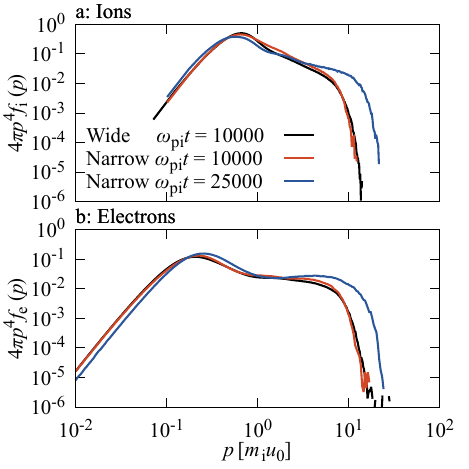}
\caption{Downstream particle momentum spectrum for $u_0/c=4/3$.
         The format is the same as in the top two panels of Figure~\ref{fig:spectrum}.
         Black lines are from the wide box simulation, used in the main text, at $\omega_{\mathrm{pi}}t=10000$.
         Red and blue lines correspond to the narrow box run at $\omega_{\mathrm{pi}}t=10000$ and $25000$, respectively.}
\label{fig:longterm}
\end{figure}

In the main text, we show that high-velocity shocks $(u_0/c\gtrsim1)$ do not transition to a Bell-dominated regime during the course of our simulation.
We ran 1D periodic box simulations in Appendix \ref{app:twocompbell} and argued that this is a physical, persistent effect due to the presence of relativistic beam electrons in high-velocity shocks, and not due to the limited timespan of our simulations.
Here, we show simulation results of the $u_0/c=4/3$ case up to $\omega_{\mathrm{pi}}t=25000$, to further support the persistence of the Weibel-dominated state at high velocities.
We use a narrower box with $L_y=10.24\,d_{\mathrm{i}}$, which is still wide enough to include a couple of Weibel filaments (see Figure~\ref{fig:shock}c2).
Figure~\ref{fig:longterm} shows the downstream particle spectra, all measured in the same region as in the main text $(x-x_{\mathrm{sh}})/d_{\mathrm{i}}=[-100, -50]$.
The red curves, which correspond to $\omega_{\mathrm{pi}}t=10000$, are consistent with the results of the wide box (in black) for both ions (panel a) and electrons (panel b).
The blue curves show the downstream spectra at $\omega_{\mathrm{pi}}t=25000$ for the narrower box.
The thermal population is mostly unchanged as compared to earlier times, while the nonthermal tail is more extended for both ions (panel a) and electrons (panel b).
We confirmed that the structure of the flow is consistent between the wide box and the narrow box at $\omega_{\mathrm{pi}}t=10000$, apart from the size of the largest density cavities in the near upstream (Figure~\ref{fig:shock}c1, $(x-x_{\mathrm{sh}})/d_{\mathrm{i}}=[0, 50]$), which is limited by the box width in the narrow box case.
In the narrow box run, the magnetic field structure is qualitatively the same at $\omega_{\mathrm{pi}}t=10000$ and $25000$ (not shown).
We argue that this shock would remain Weibel-dominated, and electrons would continue to be accelerated.
Recall that these results are complementary to and further support the conclusions of Appendix \ref{app:twocompbell}.

\bibliography{main}{}
\bibliographystyle{aasjournalv7.1}

\end{CJK*}
\end{document}